\pdfoutput=1
\documentclass[runningheads]{llncs}
\usepackage[T1,T5]{fontenc}
\usepackage[utf8]{inputenc}
\usepackage[vietnamese,english]{babel}
\usepackage{subcaption}
\usepackage{minted}
\usepackage{svg}
\usepackage{amsmath}
\usepackage{etoolbox}  
\usepackage{graphicx}
\usepackage[misc]{ifsym}

\usepackage{listings}
\usepackage{xcolor}
\usepackage{tcolorbox}

\newcommand{\equalcontrib}{\textsuperscript{*}}

\begin{document}
\title{MERVIN: A Unified Framework for Multimodal Event Retrieval in Vietnamese News Videos}

\author{
  Anh-Tai Pham-Nguyen\equalcontrib\textsuperscript{(\Letter)}\inst{1,2}\orcidID{0009-0004-1701-8976} \and
  Tung-Duong Le-Duc\equalcontrib\textsuperscript{(\Letter)}\inst{1,2}\orcidID{0009-0001-2603-0908} \and
  Anh-Duy Le\equalcontrib\textsuperscript{(\Letter)}\inst{1,2}\orcidID{0009-0008-3791-3598} \and
  Trung-Hieu Truong-Le\textsuperscript{(\Letter)}\inst{1,2}\orcidID{0009-0002-6615-171X}
}

\institute{
  University of Science, Ho Chi Minh City, Vietnam \and
  Vietnam National University - Ho Chi Minh City, Vietnam\\
  \email{pnatai23@apcs.fitus.edu.vn, 23125081@student.hcmus.edu.vn, laduy232@clc.fitus.edu.vn, tlthieu2428@clc.fitus.edu.vn}
}
\maketitle

\begingroup
  \renewcommand{\thefootnote}{\fnsymbol{footnote}}
  \setcounter{footnote}{1}
  \footnotetext[1]{These authors contributed equally to this work.}
\endgroup

\begin{abstract}

The growth of online video platforms drives the need for effective, semantically grounded event retrieval. We present MERVIN, a unified multimodal framework for Vietnamese news videos that integrates keyframes, transcripts, and video summaries. Transcript quality is enhanced via Gemini 1.5 Flash, reducing noise from accents, background sounds, and recognition errors. Visual features are extracted with Perception Encoder, while a Vietnamese language model produces textual embeddings; both are indexed in Milvus for efficient similarity-based retrieval. In addition, a React-based interface enables iterative query refinement across modalities, improving semantic alignment. Experimental results on Vietnamese news videos demonstrate the effectiveness of the proposed system, with MERVIN achieving 79 out of 88 points in \textit{AI Challenge HCMC 2025} qualification phase and successfully retrieved all results for every query in the final round.

\end{abstract}

\vspace{-0.9cm}
\section{Introduction}

The rapid growth of online video platforms has introduced substantial challenges for effective information retrieval. Ad-hoc Video Event Retrieval (VER)---the task of locating relevant video segments based on natural language queries---is a crucial capability for navigating large-scale video repositories. This challenge is exemplified by the annual \textit{Video Browser Showdown} (VBS) and \textit{Lifelog Search Challenge} (LSC), which emphasizes fast, interactive, and semantically grounded search under strict time constraints.

Two primary issues make this task particularly demanding. First, video data is inherently multimodal: meaning arises from the joint interaction of visual, auditory, and textual cues. Reliance on a single modality often leads to incomplete understanding. Second, automatically generated transcripts are prone to recognition errors, especially in non-English languages. For Vietnamese, automatic speech recognition (ASR) systems are highly susceptible to noise from background interference, speaker accents, and domain-specific vocabulary, which can significantly reduce retrieval accuracy when relying on textual features.

To mitigate these challenges, we present MERVIN---an interactive multimodal retrieval system tailored for Vietnamese news videos, developed by our team, chmod, for the \textit{AI Challenge HCMC 2025}. MERVIN integrates established visual and textual encoders within a unified framework, supported by a Large Language Model (LLM)-based preprocessing pipeline that improves the quality of noisy transcripts. Specifically, the Gemini Flash 1.5 API is employed to clean, normalize, and summarize ASR-generated text, enhancing its suitability for downstream retrieval tasks.

Feature extraction is organized into two parallel streams. The visual stream employs the PE-Core-bigG-14-448 model for keyframe embedding, while the textual stream uses the Vietnamese-specific model dangvantuan/vietnamese-embedding to represent cleaned and summarized transcripts. Both modalities are stored and indexed in a Milvus vector database to enable efficient similarity-based retrieval. A web-based interface developed in React provides user access to frame-, transcript-, and summary-level search, supporting iterative query refinement and verification.

The main contributions of this work are as follows:
\begin{enumerate}
    \item An end-to-end multimodal video retrieval system integrating visual embeddings with Vietnamese text encoders;
    \item A lightweight LLM-based pipeline for transcript cleaning and summarization to reduce ASR noise;
    \item An interactive, task-oriented interface designed to facilitate efficient event retrieval and validation.
\end{enumerate}

The remainder of this paper describes the system architecture, implementation details, and evaluation results on Vietnamese news video datasets.

\vspace{-0.2cm}
\section{Related Work}
\vspace{-0.2cm}
\subsubsection{Multimodal Vision-Language Models}
Contrastive Language-Image Pretraining (CLIP) \cite{radford2021learning} aligns text and images via contrastive learning on large datasets, enabling zero-shot classification and retrieval. OpenCLIP \cite{ilharco2021openclip} extends this framework with open-source implementations and large-scale training (e.g., LAION-5B \cite{schuhmann2022laion5b}). Variants like SigLIP \cite{zhai2023siglip} enhance multilingual understanding and fine-grained grounding, while improvements in positional encoding \cite{shaw2018self} allow longer, context-rich text processing.

\vspace{-0.2cm}
\subsubsection{Perception Encoder (PE)}
Meta's Perception Encoder (PE) family \cite{deit2024perceptionencoder} advances visual embeddings for images and videos via unified vision-language objectives, achieving 85.4\% top-1 on ImageNet-1K and 76.9\% on Kinetics-400. PE-Core-bigG-14-448 excels in zero-shot text-to-image/video retrieval, using contrastive pretraining and synthetic video data for robust, transferable embeddings applicable to retrieval, detection, tracking, and multimodal question answering.

\vspace{-0.2cm}
\subsubsection{Text Understanding Models}
Transformer-based models such as BERT \cite{devlin2019bert} and GPT-4 \cite{openai2023gpt4} remain foundational for semantic text representation. Integrated into multimodal frameworks like SigLIP, they enhance region-based captioning and multilingual comprehension, with advanced positional encoding supporting longer sequences.

\vspace{-0.2cm}
\subsubsection{Video Retrieval}

Modern video retrieval emphasizes cross-modal alignment to improve relevance and efficiency. CLIP4Clip \cite{luo2022clip4clip} adapts CLIP for video-text retrieval using mean pooling and self-attention, achieving state-of-the-art performance on MSR-VTT and ActivityNet. Frozen in Time \cite{bain2021frozen} jointly trains video and text encoders without keyframe extraction. MIL-NCE \cite{miech2020endtoend} applies multiple-instance learning for fine-grained temporal matching. These approaches illustrate a trend toward unified, context-aware retrieval pipelines. Hybrid systems integrating CLIP-based and transformer models remain effective for multilingual and real-time datasets such as the 2025 HCMC AI Challenge \cite{AIChallenge2025}.

Building on these efforts, prior work~\cite{10.1007/978-981-96-4291-5_19} demonstrates effective multimodal Vietnamese news retrieval, combining CLIP ViT-L/14, TASK-former, transcript, and OCR-based methods to handle multilingual, event-based queries in the 2024 HCMC AI Challenge. The system adapts CLIP-based techniques to large-scale, time-sensitive videos, emphasizing multimodal fusion, efficient keyframe indexing, and practical user-oriented retrieval interfaces.

\vspace{-0.2cm}
\section{Proposed Approach}

\begin{figure}[h!]
  \centering
  \includegraphics[width=.9\textwidth]{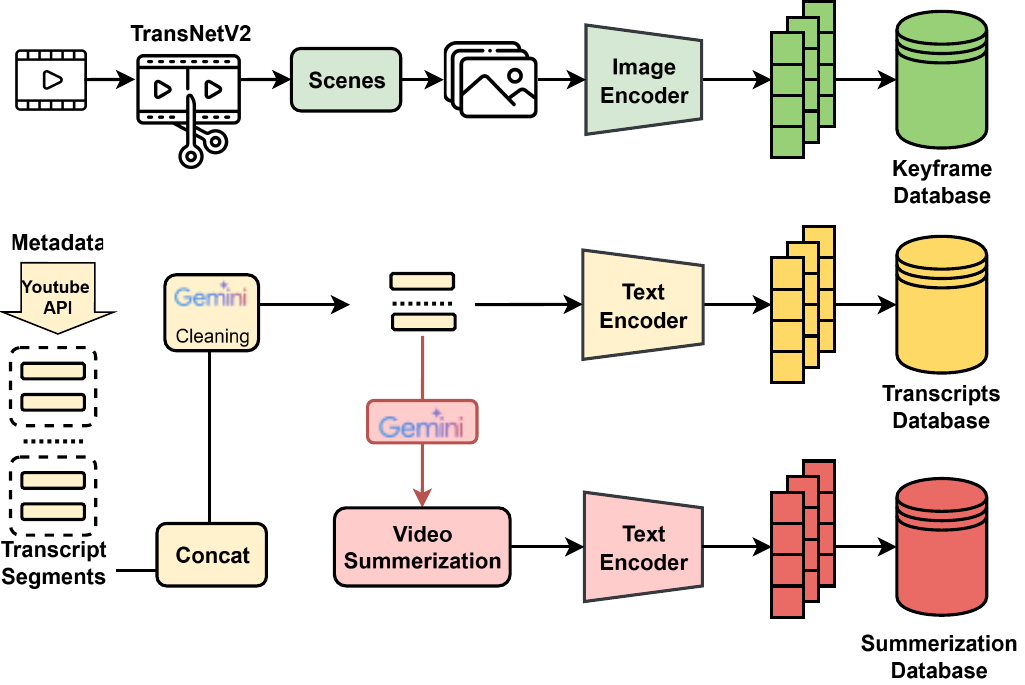}
  \caption{Overview of the data preparation pipeline of MERVIN.}
  \label{fig:data-preparation}
\end{figure}

Figure~\ref{fig:data-preparation} shows MERVIN's data preparation pipeline. Preprocessing extracts keyframes, processes transcripts, and summarizes videos, while feature extraction converts these into visual and textual embeddings for retrieval.

\vspace{-0.2cm}
\subsection{Data Preprocessing}
\vspace{-0.2cm}
\subsubsection{Keyframe Extraction}

TransNetV2~\cite{DBLP:journals/corr/abs-2008-04838} detects shot boundaries, segmenting videos into distinct shots. From each shot, three keyframes are extracted at normalized positions 0.15, 0.50, and 0.85, providing a balanced temporal representation.

\vspace{-0.2cm}
\subsubsection{Transcript Extraction}
Transcripts are obtained via the \textit{YouTube Transcript API}\footnotemark[2]; if unavailable or the video is private, audio is transcribed using OpenAI Whisper~\cite{radford2022robustspeechrecognitionlargescale}. Each transcript is split into short, fixed-length segments, with \(k=5\) consecutive segments grouped into a single interval to preserve event-level semantics. 

To improve retrieval accuracy, all segments undergo text cleaning and summarization via the Gemini Flash 1.5 API. Cleaning removes unrecognized tokens, normalizes accent variations, and resolves contextually ambiguous phrases, while summarization produces concise event-level representations. This dual-stage processing reduces noise in textual features, enabling more accurate semantic search. On average, cleaning consumes approximately 8,000 tokens and summarization 3,000 – 4,000 tokens.

\vspace{-0.2cm}
\subsubsection{Video Summerization}

From the cleaned and concatenated transcripts, event summaries for each video are generated using Gemini 1.5 Flash, highlighting key topics and major events.

\vspace{-0.2cm}
\subsection{Feature Extraction}
\vspace{-0.2cm}
\subsubsection{Keyframe}

To reduce computation and speed up queries, we use PE-Core-bigG-14-448 from Meta's PE-Core family to extract keyframe embeddings. It was chosen for its strong zero-shot performance (85.4\% ImageNet-1K, 58.1\% COCO text-to-image, 51.2\% VTT), superior to CLIP ViT-H/14 and OpenCLIP ViT-bigG/14. Its video pretraining with synthetic temporal data improves motion-consistent representations for shot-level retrieval, while attention pooling, high-dimensional embeddings, and longer text context (72 tokens) enable fine-grained alignment of visual and textual modalities, enhancing early-stage recall efficiently.

\vspace{-0.2cm}
\subsubsection{Text}
We obtain embeddings for each text segment using the pre-trained dangvantuan / vietnamese-embedding\footnotemark[3], selected based on its strong performance on the Vietnamese Semantic Textual Similarity (STS) benchmark \cite{huggingface:dataset:stsb_multi_mt} (Table~\ref{tab:embedding-benchmark}). All transcript and summary embeddings are stored in separate Milvus databases for efficient retrieval.

\vspace{-0.2cm}

\begin{table}[h!]
\centering
\caption{Performance of Vietnamese embedding models on the STS benchmark.}
\label{tab:embedding-benchmark}
\begin{tabular}{lcc}
\hline
\textbf{Model} & \textbf{Pearson (cosine)} & \textbf{Spearman (cosine)} \\
\hline
dangvantuan/vietnamese-embedding\footnotemark[3] & 0.8502 & 0.8499 \\
thanhtantran/Vietnamese\_Embedding\_v2\footnotemark[4] & 0.7565 & 0.7587 \\
thanhtantran/Vietnamese\_Reranker\footnotemark[5] & 0.1558 & 0.1577 \\
Alibaba-NLP/gte-Qwen2-1.5B-instruct\footnotemark[6] & 0.4418 & 0.5547 \\
avsolatorio/GIST-Embedding-v0\footnotemark[7] & 0.5565 & 0.5602 \\
\hline
\end{tabular}
\end{table}

\vspace{-0.2cm}
\subsection{Interactive User Interface}

\begin{figure}
    \centering
    \includegraphics[width=\linewidth]{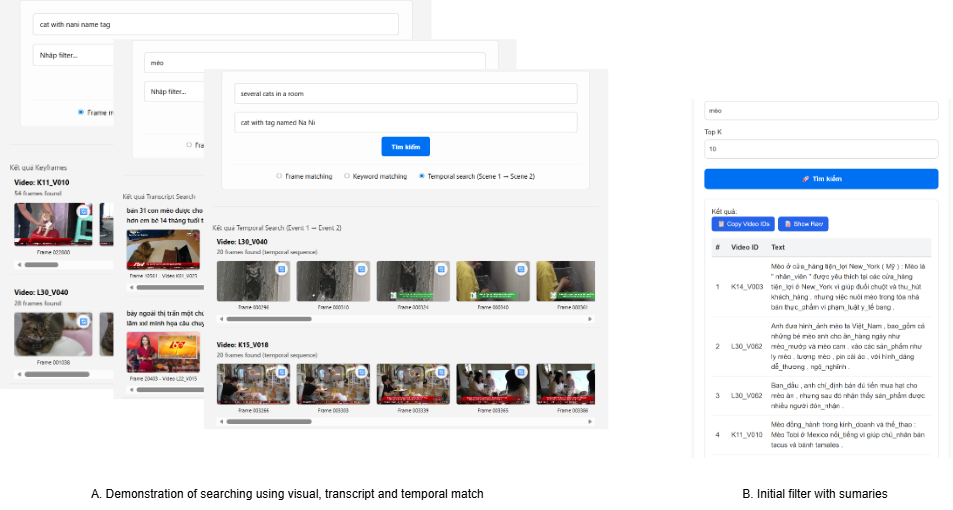}
    \caption{Demonstration of our user interface: (A) searching with visual and transcript matching, and (B) displaying initial retrieved videos with summaries for filtering. }
    \label{fig:ui}
\end{figure}

To facilitate interaction with the proposed event retrieval system, we built a web interface with four modules (Figure~\ref{fig:ui}) using \textit{React}, providing a responsive, asynchronous experience for seamless query refinement.

\vspace{-0.5cm}
\subsubsection{Frame-, Transcript-, and Temporal-Based Search}
Frame- and transcript-based search take a single text query, while temporal search requires two events (E1 and E2) to retrieve a valid sequence. All methods return relevant frame identifiers grouped by source video, which are then ranked for the final video-level display. Selecting a frame plays the corresponding segment via the \textit{YouTube Player API} (React), dynamically showing the current frame index computed from the video's FPS.

\footnotetext[3]{\url{https://huggingface.co/dangvantuan/vietnamese-embedding}}
\footnotetext[4]{\url{https://huggingface.co/thanhtantran/Vietnamese\_Embedding\_v2}}
\footnotetext[5]{\url{https://huggingface.co/thanhtantran/Vietnamese\_Reranker}}
\footnotetext[6]{\url{https://huggingface.co/Alibaba-NLP/gte-Qwen2-1.5B-instruct}}
\footnotetext[7]{\url{https://huggingface.co/avsolatorio/GIST-Embedding-v0}}

\newpage

\vspace{-0.2cm}
\subsubsection{Summarization-Based Search}
Summarization-based search supports higher-level semantic retrieval, returning video IDs and summarized transcripts without frame-level matches. Users can verify candidate videos and convert timestamps to frame indices as needed, complementing frame-level retrieval for multi-scene or event-level queries.

\vspace{-0.2cm}
\subsubsection{Efficient Query Handling and System Responsiveness}
To maintain low latency, videos are streamed directly from YouTube rather than hosted locally, reducing storage and improving responsiveness. Frame positions are computed on-the-fly from playback time and FPS, providing near-instant feedback even during rapid query updates.

\vspace{-0.2cm}
\subsubsection{Submission and Verification Page}
\begin{figure}
    \centering
    \includegraphics[width=0.9\linewidth]{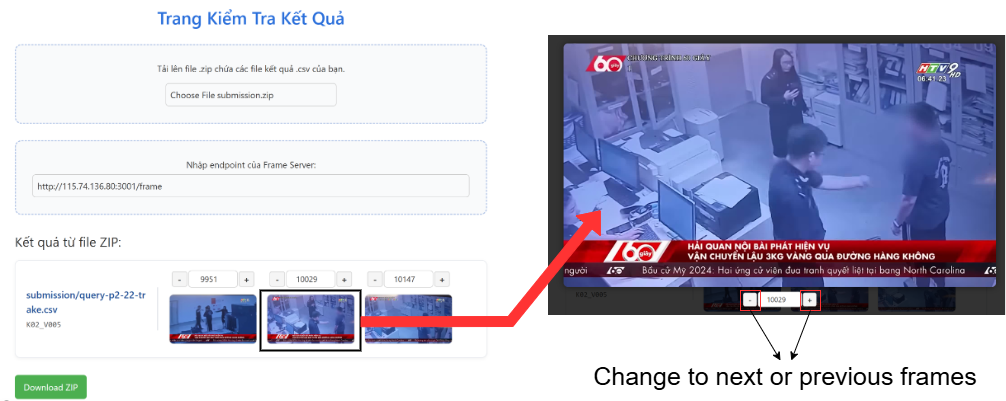}
    \caption{Demonstration of Submission and Verification Page}
    \label{fig:submission}
\end{figure}

The submission page (Figure~\ref{fig:submission}) allows users to review retrieved frames before submission. \textit{FFmpeg} extracts frames directly from video files and iteratively validate the correspondence between frame indices and timestamps. After verification, the submission archive (ZIP) is automatically updated with only confirmed frames.

\vspace{-0.5cm}

\subsection{Query}

\begin{figure}[h!]
  \centering
  \includegraphics[width=.9\textwidth]{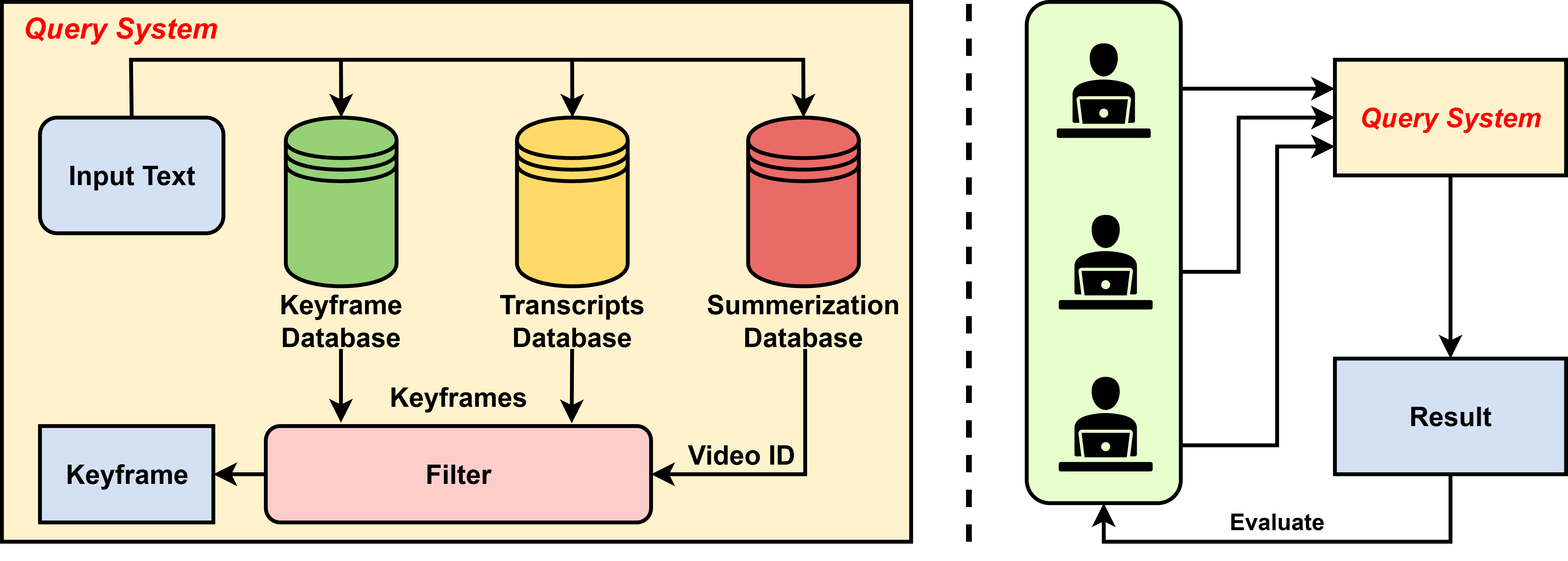}
  \caption{Overview of the query pipeline of MERVIN.}
  \label{fig:query}
\end{figure}

Figure~\ref{fig:query} illustrates the query process. Users first identify the most informative terms, which guide retrieval and filtering across three databases: keyframes, transcripts, and summaries.

\vspace{-0.4cm}
\subsubsection{Keyframe Extraction}
The query retrieves the top-$k$ keyframes ($k=1000$). Each keyframe is manually inspected, with video playback support provided for context verification.

\vspace{-0.2cm}
\subsubsection{Transcript Extraction}
The query retrieves the top-$k$ transcript segments, showing all keyframes within each segment. Additionally, if the original query contains distinctive or domain-specific terms, these can be used to further refine and filter the retrieved transcripts.

\vspace{-0.2cm}
\subsubsection{Advanced Search}
The processed query is expanded to locate semantically related videos via stored summaries, filtering and prioritizing previously retrieved keyframes to improve coherence and alignment with the query intent.

\vspace{-0.2cm}
\subsubsection{Iterative Refinement and evaluation}

Users assess results, refine the query, reprocess it and  fed back into the \textit{Query System} iteratively. Through this cycle, user efficiently locates the video segments and keyframes corresponding to the intended event.

\vspace{-0.2cm}
\subsubsection{Temporal Search}
To identify a sequence of two events, $E_1$ and $E_2$, we first execute independent queries for each. The resulting event pairs are then subjected to temporal filtering. We prune any candidates where the timestamp for the second event ($T_2$) precedes the first ($T_1$), or where the duration ($T_2 - T_1$) exceeds a five-minute threshold.

Remaining candidates are ranked using a heuristic score ($S_{\text{video}}$). This score prioritizes the best-detected sequential pair ($S_{\text{pair}}$) but also considers the average top-10 scores for the individual events ($\bar{S}_1, \bar{S}_2$):
\begin{equation}
  S_{\text{video}} = (10.0 \times S_{\text{pair}}) + (5.0 \times (\bar{S}_1 + \bar{S}_2))
  \label{eq:temporal_score}
\end{equation}

\vspace{-0.8cm}
\section{Evaluation}
\vspace{-0.2cm}
\subsection{Experimental Setup}



The system was deployed following the AI Challenge HCMC 2025 \cite{AIChallenge2025} infrastructure. The backend, handling text and video embeddings, ran on a machine with an \textbf{AMD Ryzen 5 5600G} CPU (12 cores, 3.9\,GHz), \textbf{32\,GB RAM}, and \textbf{NVIDIA RTX 3060} GPU (12\,GB VRAM), with stable network performance (309\,Mbps download, 317\,Mbps upload, 2\,ms latency). Frame visualization and streaming were handled by a separate machine, while users operated individual clients for retrieval, simulating realistic concurrent usage.

\vspace{-0.2cm}
\subsection{Evaluation Protocol}
System performance was evaluated in two distinct phases as defined by the \textit{AI Challenge HCMC 2025}. In the \textbf{qualification phase}, teams submitted up to 100 ranked results per query. Performance was measured by the \textbf{Mean of Top-k R-Scores} (using $k \in \{1, 5, 20, 50, 100\}$), computed as:
\[
\text{Final Score} = \frac{1}{5} \sum_{k} \max_{1 \le i \le k} \text{R-Score}(r_i)
\]
where $r_i$ is the $i$-th result. The R-Score quantified correctness: for \textbf{KIS} tasks, the video and frame index had to match; for \textbf{VQA}, the video, frame, and answer had to match; and for \textbf{Temporal Alignment}, the segment had to overlap the ground truth.

For the \textbf{final round}, we report on the number of queries our team successfully answered within the time constraints.

\vspace{-0.2cm}
\subsection{Illustrative Example}

To demonstrate the effectiveness of frame-based and temporal search, we use \textit{tkis-query-10} as an example:

\begin{tcolorbox}[colback=gray!10,colframe=black]
Vietnamese:
Năm 2019, ba nhà khoa học từ Mỹ, Anh và Nhật Bản nhận giải Nobel Hóa học cho công trình phát triển một vật dụng quen thuộc. Tìm clip về hoạt động tại Úc nhằm thu gom và tái chế vật dụng này và các vật dụng tương tự. Video có thùng thu gom hình chữ nhật, nắp xanh dương, viền xanh lá, với ký hiệu hướng dẫn xung quanh, đặt sát tường màu gỗ. Cuối clip xuất hiện cửa hàng Coles màu trắng trên nền đỏ.

\medskip

English:
In 2019, three scientists from the US, UK, and Japan were awarded the Nobel Prize for developing a familiar everyday item. Find the clip showing an activity in Australia aimed at collecting and recycling this item and similar ones.
\end{tcolorbox}

\begin{figure}[H]
    \centering
    \begin{subfigure}[b]{0.40\linewidth}
        \centering
        \includegraphics[width=\linewidth]{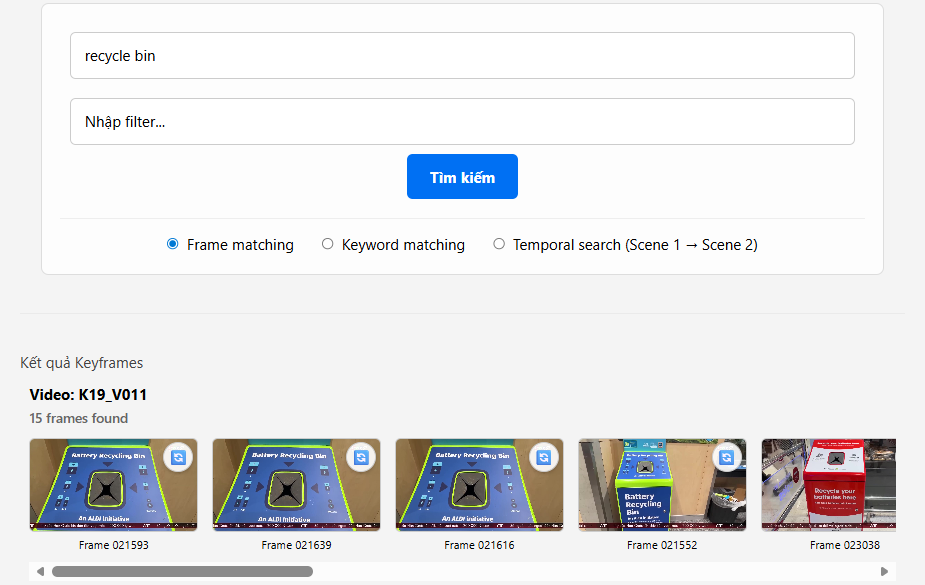}
        \caption{Frame search for \textit{tkis-query-10}.}
        \label{fig:kis-frame}
    \end{subfigure}
    \hfill
    \begin{subfigure}[b]{0.48\linewidth}
        \centering
        \includegraphics[width=\linewidth]{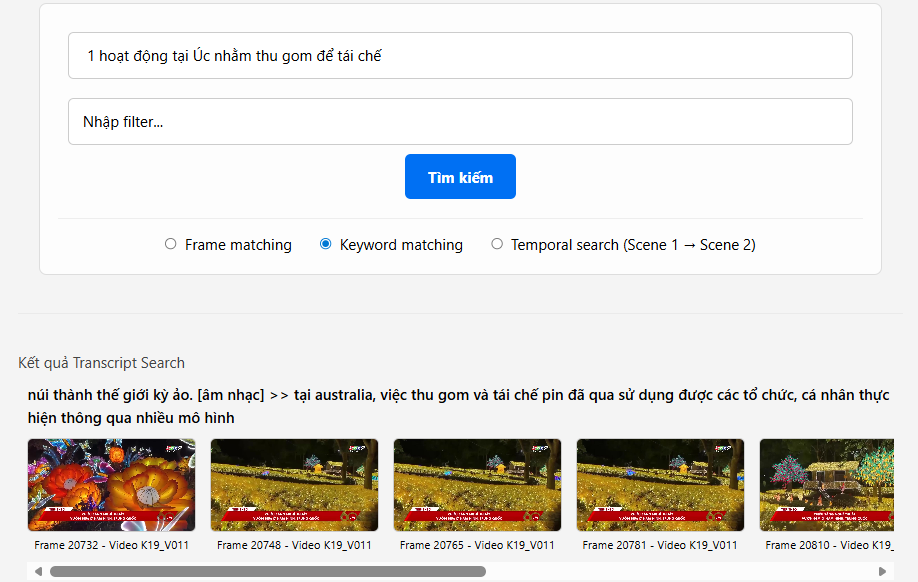}
        \caption{Transcript search for \textit{tkis-query-10}.}
        \label{fig:kis-transcript}
    \end{subfigure}

    \vspace{0.3em} 
    \begin{subfigure}[b]{0.48\linewidth}
        \centering
        \includegraphics[width=\linewidth]{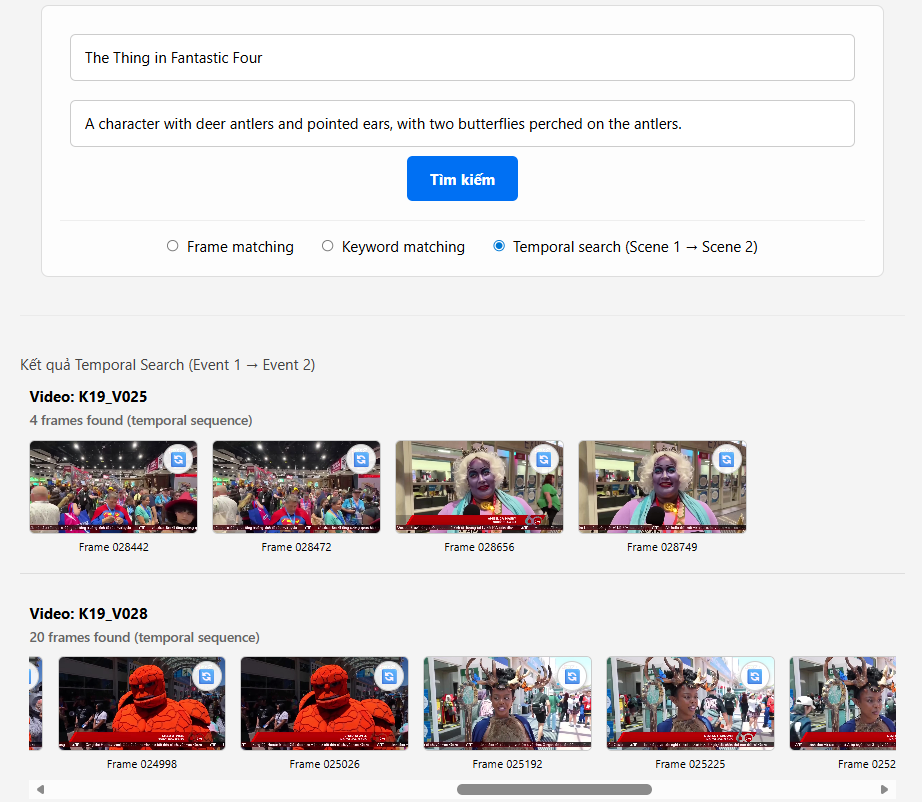}
        \caption{Temporal search for \textit{trake-01}, showing frames for events E1 and E2.}
        \label{fig:trake-temporal}
    \end{subfigure}

    \caption{Comparison of search results for \textit{tkis-query-10} and \textit{trake-01}.}
    \label{fig:search-results}
\end{figure}

This clip illustrates the collection and recycling of batteries (including lithium-ion) by the Aldi supermarket chain in Australia. The results are shown in Figures~\ref{fig:kis-frame} and~\ref{fig:kis-transcript}, both returning the relevant video as the top-1 result.

To illustrate temporal search, we use the \textit{trake-01} query. The task is as follows.

\begin{lstlisting}
At a festival with many people in costume, identify the first moment in the video when the following costumed characters appear prominently in the frame:

    E1: The Thing from Fantastic Four (character with rock-like skin)
    E2: A character with deer antlers and pointed ears, with two butterflies perched on the antlers
    E3: Wonder Woman
    E4: Greta, a female Gremlin with green skin and large ears
\end{lstlisting}

As shown in Figure~\ref{fig:trake-temporal}, the temporal search using E1 and E2 returns the target video within the top-2 results, with the required frames correctly identified for each event.

\vspace{-0.2cm}
\subsection{Results}

Our system was evaluated in the \textit{AI Challenge HCMC 2025}. During the qualification phase, we attained 79 out of 88 possible points, as summarized in Table~\ref{tab:qualification_results}. This performance ranked us among the top university teams advancing to the final round.

In the subsequent final round, our system successfully retrieved all desired scenes, confirming its effectiveness in the live competition setting.

\vspace{-0.5em}

\begin{table}
\caption{Qualification phase results and official scoring scheme for Track A (University Division).}
\label{tab:qualification_results}
\centering
\begin{tabular}{|l|c|c|c|c|}
\hline
\textbf{Round} & \textbf{Round 1} & \textbf{Round 2} & \textbf{Round 3} & \textbf{Total Score} \\
\hline
\textbf{Score Achieved} & 18 & 27 & 34 & \textbf{79} \\
\textbf{Maximum Score} & 23 & 30 & 35 & \textbf{88} \\
\hline
\end{tabular}
\end{table}

\vspace{-1cm}
\subsection{Discussion}

In the qualification phase, MERVIN demonstrated competitive and stable retrieval performance, attaining 79 out of 88 points and showing adaptability across different query sets. Notably, the system effectively retrieved target videos and keyframes even when provided with only partial keyword queries.

Performance was particularly strong for KIS-type queries, especially those containing rich visual descriptions. This efficacy was supported by the system's distributed architecture, which decouples embedding computation from frame visualization and user interaction. This design proved crucial for maintaining low latency and enabling smooth real-time retrieval during live testing.

\vspace{-0.2cm}
\section{Future Work}
Building on the framework presented, this work highlights two promising directions for future research in interactive video retrieval. First, domain-specific ASR error correction models could be investigated to improve the reliability of textual features. Second, a significant opportunity lies in exploring the use of LLM agents to automate parts of the retrieval workflow. Such agents could autonomously manage tasks currently requiring manual intervention, like query refinement and result verification, thereby increasing overall system efficiency.

\vspace{-0.2cm}
\section{Conclusion}
This paper presents MERVIN, a unified multimodal framework for Vietnamese video event retrieval that integrates visual and textual modalities through an end-to-end pipeline. By combining PE-Core-bigG-14-448 for keyframe representation with Gemini 1.5 Flash for transcript cleaning and summarization, MERVIN effectively reduces ASR noise and enhances cross-modal alignment.
Its interactive query interface enables iterative refinement across keyframes, transcripts, and summaries, allowing users to precisely localize video events. Evaluated at the \textit{AI Challenge HCMC 2025}, the system demonstrated strong performance, scoring 79 out of 88 points and ranking among the top university teams.

This work highlights promising directions for future research in interactive video retrieval. Enhancements could include domain-specific ASR error correction models to improve textual feature reliability. Furthermore, a significant opportunity lies in exploring the use of LLM agents to automate parts of the retrieval workflow, such as query refinement and result verification, to increase overall system efficiency.

\section*{Acknowledgments}
This research is supported by research funding from the Faculty of Information Technology,
University of Science, Vietnam National University - Ho Chi Minh City.

\bibliographystyle{splncs04}

\bibliography{main.bbl}

\end{document}